\documentclass[12pt,a4paper]{article}
\usepackage[textheight=23cm,textwidth=17.5cm]{geometry}
\usepackage{amsmath}
\usepackage{graphicx}
\usepackage{cite}
\usepackage[american]{babel}
\usepackage{amssymb}
\usepackage{verbatim}
\usepackage{textcomp}
\usepackage{threeparttable}

\usepackage{endfloat}
\AtBeginDelayedFloats{\linespread{1.3}}

\usepackage{dcolumn} 
\newcolumntype{d}{D{.}{.}{2}}
\newcolumntype{e}{D{.}{.}{3}}
\newcolumntype{f}{D{.}{.}{4}}
\newcommand{\fett}[1]{\mbox{\boldmath$#1$}}

\newcommand{\bra}[1]{\ensuremath{\left\langle #1\right|}}
\newcommand{\ket}[1]{\ensuremath{\left|#1\right\rangle}}
\newcommand{\braket}[2]{\ensuremath{\left\langle #1\vphantom{#2}\right.\left|\vphantom{#1}#2\right\rangle}}

\newcommand{\tr}[0]{\ensuremath{\mathrm{Tr}}}
\DeclareMathAlphabet{\mathpzc}{OT1}{pzc}{m}{it}
\DeclareMathSizes{12}{13}{7}{8}

\def\trs{^\mr{T}}
\newcommand{\mx}[1]{\boldsymbol{#1}}
\newcommand{\mr}[1]{\mathrm{#1}}
\def\Eh{E$_\mr{h}$}
\usepackage{color}

\begin{document}

\begin{center}
~\vspace*{-0.01cm}
{\Large %
  Elimination of the Translational Kinetic Energy Contamination in pre-Born--Oppenheimer Calculations %
}\\
\vspace{0.3cm}
{\large
Benjamin Simmen$^a$, Edit M\'atyus$^b$\footnote{corresponding author; e-mail: matyus@chem.elte.hu}, Markus Reiher$^a$\footnote{corresponding author; e-mail: markus.reiher@phys.chem.ethz.ch}
}\\[2ex]

$^a$ETH Zurich, Laboratorium f\"ur Physikalische Chemie, \\
Wolfgang-Pauli-Str.~10, 8093 Zurich, Switzerland\\
$^b$Laboratory of Molecular Structure and Dynamics, Institute of Chemistry, E\"otv\"os University\\
 P\'azm\'any P\'eter s\'et\'any 1/A, H-1117 Budapest, Hungary

December 15, 2012 \\[-0.5cm]

\end{center}

\abstract
In this paper we present a simple strategy for the elimination of the translational kinetic energy contamination of the total energy in pre-Born--Oppenheimer calculations carried out in laboratory-fixed Cartesian coordinates (LFCCs).
The simple expressions for the coordinates and the operators are thus preserved throughout the calculations,
while the mathematical form and the parametrisation of the basis functions are chosen so that the translational and
rotational invariances are respected. The basis functions are constructed using
explicitly correlated Gaussian functions (ECGs) and the global vector representation.

First, we observe that it is not possible to parametrise the ECGs so that the system is at rest in LFCCs
and at the same time the basis functions are square-integrable with a non-vanishing norm.
Then, we work out a practical strategy to circumvent this problem by making use of the properties of the linear transformation between the LFCCs and translationally invariant and center-of-mass Cartesian coordinates
as well as
the transformation properties of the corresponding basis function parameter matrices.
By exploiting these formal mathematical relationships we can identify and
separate the translational contamination terms in the matrix representation of
the kinetic energy operator in the LFCC formalism.

We present numerical examples for the translational contamination and its elimination for
the two lowest rotational energy levels of the singlet hydrogen molecule, corresponding to para- and ortho-H$_2$, respectively, treated as four-particle quantum systems.

\newpage
\normalsize
\section{Introduction}

Expressions for the calculation of intrinsic properties of molecules should be free of contributions from the overall translation of the system. In the commonly introduced Born--Oppenheimer (BO) approximation, the nuclei are fixed, and thus the translational contribution is automatically separated. In several combined electron-nuclear orbital approaches 
\cite{Chakraborty2008,Bochevarov2004,Goli2012} the translational dependence is eliminated automatically by fixing one or a few heavy particles. Here, we consider molecules as many-particle quantum systems with electrons and nuclei both treated as quantum particles on equal footing in the pre-Born--Oppenheimer (pre-BO) quantum theory.

Traditionally, in rovibrational calculations, in which all nuclei are treated as quantum particles on a potential energy surface 
\cite{Albert2006,Albert2011,Luckhaus2000,Kuhn1999,Quack2001}, the first step is the separation of the Cartesian coordinates of the center of mass followed by the definition of a body-fixed frame, orientational angles, and internal coordinates. This approach results in the replacement of the original laboratory-fixed Cartesian coordinates with curvilinear coordinates and the corresponding very complicated, translationally invariant rotational-vibrational Hamiltonians, see for example \cite{Matyus2009}.

Less complicated translationally invariant Hamiltonians are used in full pre-BO calculations, where the original laboratory-fixed Cartesian coordinates are replaced by some set of translationally invariant Cartesian coordinates (TICCs) and the center-of-mass Cartesian coordinates (CMCCs) are separated 
\cite{Cafiero2003,Bubin2012,Matyus2011,Matyus2011a,Matyus2012}. Although the resulting TICCs are rectilinear coordinates, the corresponding Hamiltonian is still complicated. It is therefore reasonable to ask whether we can make our calculations even simpler, using the original laboratory-fixed-Cartesian-coordinate (LFCC) formalism without having to rely on any coordinate transformation at all. In this work we therefore explore the usage of LFCCs in pre-BO calculations.

If LFCCs are used one has to make sure that the energy of the overall translation of the system is eliminated.
The most straightforward way is the subtraction of the kinetic energy operator of the center of mass from the total Hamiltonian, which, however, requires the evaluation of an additional integral with mixed coordinate second derivatives \cite{Nakai2002,Nakai2005}. To avoid this additional integral evaluation we develop here an alternative approach.

In short, our computational strategy in the LFCC formalism to obtain eigenstates with various angular momentum quantum number is as follows. In our variational procedure we use basis functions, which are eigenfunctions of the total spatial angular momentum operators, $\hat{L}^2$ and $\hat{L}_z$ \cite{Matyus2012}. This is the simplest way to make sure that we obtain angular momentum eigenstates, since rotational ``contamination'' cannot be removed by a simple subtraction of a term from the full Hamiltonian \cite{Nakai2005,Sutcliffe2005}. Then, we investigate the effect of the parametrisation of the basis functions on the translational contamination of the total energy and correct for it during the evaluation of the integrals in the LFCC formalism.

The paper is organised as follows. Firstly, in Section~\ref{sec:Theo}, we present the necessary theoretical details. In Section~\ref{sec:TB}, we identify the translational energy contamination of the matrix elements. Next, in Section \ref{sec:NUM} we present numerical results for the two lowest rotational states of the H$_2$ molecule.

\section{Theoretical Details}
\label{sec:Theo}
In this section we present the theoretical details of the variational procedure applied
to solve the full quantum Hamiltonian without clamping the nuclei.
We introduce the notation essential to this work.
Firstly, we define laboratory-fixed Cartesian coordinates (LFCCs)
and translationally invariant Cartesian coordinates (TICCs) \cite{Sutcliffe2003} as well as
the corresponding non-relativistic quantum Hamiltonians. The introduction of the TICC formalism is necessary to better understand the mathematical properties of the LFCC formalism, which we then use in the calculations.
Then, we define the basis functions using explicitly correlated Gaussian functions (ECGs) \cite{Jeziorski1979,Cencek1993,Rychlewski2004,Boys1960,Singer1960}
and the global vector representation (GVR) \cite{Varga1998,Suzuki1998,Suzuki1998a}.
Finally, we present the corresponding matrix elements and point out the parametrisation problems
in the LFCC formalism.

The generalised eigenvalue problem corresponding to the matrix representation of the Hamiltonian
is solved using the standard linear algebra library routines of LAPACK (Version 3.2.1)\cite{Anderson1999} through the Armadillo framework (Version3.4.0)\cite{Sanderson2010}.

\subsection{Coordinates and the Quantum Hamiltonian}
\label{sec:H}

The non-relativistic quantum Hamiltonian for $n+1$ particles with $m_i$ masses and $q_i$ electric charges is in atomic units
\begin{align}
  \hat{H}
  =
  -\sum_{i=1}^{n+1} \frac{1}{2m_i} \Delta_{\mx{r}_i}
  +
  \sum_{i=1}^{n+1}\sum_{j>i}^{n+1} \frac{q_iq_j}{|\mx{r}_i-\mx{r}_j|}
  \label{eq:Hop}
\end{align}
where the  vector $\mx{r}=(\mx{r}_1,\mx{r}_2,\ldots,\mx{r}_{n+1})$ collects
the LFCCs.
The translational motion of the center of mass has traditionally been
separated by introducing some set of TICCs \cite{Sutcliffe2003},
$\mx{x}=(\mx{x}_1,\mx{x}_2,\ldots,\mx{x}_n)$
and the Cartesian coordinates of the center of mass (CMCC), $\mx{X}_\mathrm{CM}$,
by a linear transformation
\begin{align}
  \mx{x}_\mathrm{TICM} =
  \left[%
    \begin{array}{@{}c@{}}
      \mx{x} \\
      \mx{X}_\mr{CM} \\
    \end{array}
  \right]
  =
  (\mx{U}_x \otimes \mx{1}_3) \mx{r}
  \label{eq:defticm}
\end{align}
where $\mx{1}_3$ is the $3 \times 3$ unit matrix and
$\mx{U}_x$ is a non-singular constant matrix with the restriction
\begin{align}
  \sum_{j=1}^{n+1} (\mx{U}_x)_{ij} = 0, \quad\text{with}\quad i=1,2,\ldots,n\quad
  \label{eq:ticond1}
\end{align}
and
\begin{align}
  (\mx{U}_x)_{n+1,j} = m_j/m_\mr{tot}, \quad\text{with}\quad j=1,2,\ldots,n+1\quad ,
  \label{eq:ticond2}
\end{align}
which guarantees the translational invariance for the coordinates $\mx{x}$
and gives the definition of the center of mass.
The notation $m_\mr{tot}=\sum_{j=1}^{n+1}m_j$ was introduced.
The general form of the quantum Hamiltonian corresponding to the new coordinates $\mx{x}$ and $\mx{X}_\mr{CM}$
\cite{Sutcliffe2003,Sutcliffe2003a,Suzuki1998a} is
\begin{align}
  \hat{H}_\mr{TICM}
  =
  -\frac{1}{2}\sum_{i=1}^n \sum_{j=1}^n \mu^{(x)}_{ij} \mx{\nabla}_{\mx{x}_i}\trs \mx{\nabla}_{\mx{x}_j}
  -\frac{1}{2m_\mr{tot}} \Delta_{\mx{X}_\mr{CM}}
  +\sum_{i=1}^{n+1} \sum_{j>i}^{n+1} \frac{q_iq_j}{|(\mx{f}^{(x)}_{ij}\otimes \mx{1}_3)\trs\mx{x}|}
  \label{eq:hticm}
\end{align}
with $\mx{\nabla}_{\mx{x}_i}\trs = (\partial/\partial x_{i1},\partial/\partial x_{i2},\partial/\partial x_{i3})$,
\begin{align}
  \mu^{(x)}_{ij}
  =
  \sum_{k=1}^{n+1} (\mx{U}_x)_{ik} (\mx{U}_x)_{jk}/m_k \quad\text{with}\quad i,j=1,2,\ldots,n
\end{align}
and
\begin{align}
  (\mx{f}^{(x)}_{ij})_k
  =
  (\mx{U}_x^{-1})_{ik} - (\mx{U}_x^{-1})_{jk} \quad\text{with}\quad i,j=1,2,\ldots,n\ .
\end{align}

In Eq.~(\ref{eq:hticm}) the translational kinetic energy of the center of mass,
$\hat{T}_\mr{CM}=-1/2m_\mr{tot} \Delta_{\mx{X}_\mr{CM}}$,
separates, and by subtracting it the translationally invariant (TI) Hamiltonian,
$\hat{H}_\mr{TI}=\hat{H}-\hat{H}_\mr{CM}$, is obtained.
Mayer and Rokob \cite{Mayer2012} have recently shown that the same result can be obtained
by requiring that the total momentum of the system is zero.

Traditionally the TI form of the Hamiltonian has been used in pre-BO calculations
\cite{Cafiero2003,Bubin2012,Matyus2012}.
In this work we explore the consequences of the direct usage of the LFCC formalism and
the original Hamiltonian, Eq.~(\ref{eq:Hop}).

\subsection{Basis Functions}
\label{sec:ECG}
The basis functions are defined similarly to Ref. \cite{Matyus2012}. The spatial part is
constructed using ECGs \cite{Jeziorski1979,Cencek1993,Rychlewski2004,Boys1960,Singer1960}
and the GVR \cite{Varga1998,Suzuki1998}.
In what follows we introduce the notation relevant for this work.

The basic structure of a basis function written in LFCCs, $\mx{r}$, is
\begin{align}
  \psi_{LM_L}(\mx{r};\mx{A},\mx{u},K)
  &=
  |\mx{v}|^{2K+L}
  Y_{LM_L}(\hat{\mx{v}}) \exp\left(-\frac{1}{2} \mx{r}\trs (\mx{A}\otimes \mx{1}_3) \mx{r} \right)
  \label{eq:basdef2}
\end{align}
with the global vector
\begin{align}
  \mx{v} = (\mx{u}\otimes \mx{1}_3) \mx{r} \ .
\end{align}
The values collected in $\mx{u}=(u_1,u_2,\ldots,u_{n+1})$ are variational parameters.
Further variational parameters are $K\in\mathbb{N}_0$ and the elements of the symmetric matrix $\mx{A}$
which describe correlations between the elements of $\mx{r}$.
The matrix $\mx{A}$ must be positive definite to ensure that the basis functions
are square-integrable and have a non-vanishing norm.

Furthermore, in Eq.~(\ref{eq:basdef2}) $Y_{LM_L}(\hat{\mx{v}})$ is a spherical harmonic function
of degree $L$ and order $M_L$ and $\hat{\mx{v}}$ represents the direction of the global vector.
Thus, the parity of the basis function $\psi_{LM_L}$ is $(-1)^L$ (``natural parity'').
The quantum numbers $L$ and $M_L$ correspond to the total ``spatial'' (orbital plus rotational)
angular momentum operators, $\hat{L}^2$ and $\hat{L}_z$.

Next, we consider the transformation of the LFCCs, $\mx{r}$, to some set of TICCs, $\mx{x}$ and the CMCCs, $\mx{X}_\mr{CM}$,
in order to better understand the properties of the LFCC formalism, which we intend to use in variational
calculations.
Upon a linear transformation of the coordinates the mathematical form of the spatial functions
remains the same, and only the parameter matrices have to be transformed
\begin{align}
  \psi_{LM_L}(\mx{r};\mx{A},\mx{u},K)
  &=
  |\mx{v}|^{2K+L}
  Y_{LM_L}(\hat{\mx{v}})
  \exp\left(-\frac{1}{2} \mx{r}\trs (\mx{A}\otimes \mx{1}_3) \mx{r} \right)
  \label{eq:bstrans1} \\
  &=
  |\mx{v}|^{2K+L}
  Y_{LM_L}(\hat{\mx{v}})
  \exp\left(-\frac{1}{2} \mx{x}_\mr{TICM}\trs (\mx{A}^{(x)}\otimes \mx{1}_3) \mx{x}_\mr{TICM} \right)
  \label{eq:bstrans2}
\end{align}
with
\begin{align}
  \mx{v}
  &=
  (\mx{u}\otimes\mx{1}_3) \mx{r} \\
  &=
  (\mx{u}^{(x)}\otimes\mx{1}_3) \mx{x}_\mathrm{TICM} \ .
\end{align}
According to the definition of the $\mx{x}_\mr{TICM}$ coordinates, Eq.~(\ref{eq:defticm}),
the parameter matrices can be transformed back and forth as
\begin{align}
  \mx{A}^{(x)} = \mx{U}_x^{-\mr{T}} \mx{A} \mx{U}_x^{-1}
  \quad\Leftrightarrow\quad
  \mx{A} = \mx{U}_x^{\mr{T}} \mx{A}^{(x)} \mx{U}_x
  \label{eq:amxtrans}
\end{align}
and
\begin{align}
  \mx{u}^{(x)} = \mx{U}_x^{-\mr{T}} \mx{u}
  \quad\Leftrightarrow\quad
  \mx{u} = \mx{U}_x^{\mr{T}} \mx{u}^{(x)}
  \label{eq:uvectrans}
\end{align}

Due to the translational invariance condition, Eqs.~(\ref{eq:ticond1})--(\ref{eq:ticond2}),
the parameter matrices have block structure \cite{Matyus2012}
\begin{align}
  \mx{A}^{(x)}
  =
  \left[%
    \begin{array}{@{}cc@{}}
      \mathpzc{A}^{(x)} &   0 \\
                      0 & c_A \\
    \end{array}
  \right]
  \quad
  \text{and}
  \quad
  \mx{u}^{(x)}
  =
  \left[%
    \begin{array}{@{}c@{}}
      \mathpzc{u}^{(x)} \\
                    c_u \\
    \end{array}
  \right] \ .
  \label{eq:aublock}
\end{align}
By combining this property with Eqs.~(\ref{eq:bstrans1})--(\ref{eq:bstrans2})
we obtain
\begin{align}
  \psi_{LM_L}(\mx{r};\mx{A},\mx{u},K)
  &=
  |\mx{v}|^{2K+L}
  Y_{LM_L}(\hat{\mx{v}})
  \exp\left(-\frac{1}{2} \mx{r}\trs (\mx{A}\otimes \mx{1}_3) \mx{r} \right)
  \label{eq:bstrans1b} \\
  &=
  |\mx{v}|^{2K+L}
  Y_{LM_L}(\hat{\mx{v}})
  \exp\left(%
    -\frac{1}{2} \mx{x}\trs (\mx{\mathpzc{A}}^{(x)}\otimes \mx{1}_3) \mx{x}
    -\frac{1}{2} c_A \mx{X}_\mr{CM}^2
  \right)
  \label{eq:bstrans2b}
\end{align}
and
\begin{align}
  \mx{v}
  &=
  (\mx{u}\otimes\mx{1}_3) \mx{r} \\
  &=
  (\mx{\mathpzc{u}}^{(x)}\otimes\mx{1}_3) \mx{x}
  +
  c_u \mx{X}_\mathrm{CM} \ ,
\end{align}
which means that the system is at rest only in the case
of $c_A=0$ and $c_u=0$ \cite{Hagedorn1980,Faou2009,Matyus2012}.
At the same time we have to ensure that our earlier requirements for the
basis function parameters, $\mx{u}$ and $\mx{A}$ are fulfilled.
While the condition $c_u=0$ does not cause any problems, we find that
$c_A=0$ does so. If $c_A=0$ the matrix $\mx{A}^{(x)}$ is singular and since
$\mx{A}$ and $\mx{A}^{(x)}$ are related by a linear transformation, so is $\mx{A}$.
This violates the requirement for $\mx{A}$ to be positive definite.
Thus, we can conclude that it is not possible to parametrise ECGs in the LFCC formalism
so that the basis functions are square integrable with a non-vanishing norm and
at the same time the system is at rest.

According to the block structure, Eq.~(\ref{eq:aublock}) and the back-transformation, Eq.~(\ref{eq:amxtrans}) the exponent matrix is constructed in our calculations according to
\begin{align}
  (\mx{A})_{ij}
  =
  -\alpha_{ij}(1-\delta_{ij})
  +\left(%
    \sum_{k=1,k\neq i}^{n+1}
      \alpha_{ik}
  \right)
  \delta_{ij}
  +c_A\frac{m_i}{m_\mr{tot}}\frac{m_j}{m_\mr{tot}} \ .
  \label{eq:Amxalpha}
\end{align}

\subsection{Matrix Representation of the Hamiltonian}
\label{sec:MX}
In spite of the parametrisation difficulties described in Section~\ref{sec:ECG},
we intend to use the LFCC formalism to construct the matrix
representation for the quantum Hamiltonian because of its original simplicity.
We repeat here only the necessary expressions from Ref.~\cite{Matyus2012}
and for the original integral derivation see \cite{Suzuki1998a}.

The matrix element of the kinetic energy operator for the $I$th and $J$th
quasi-normalised basis functions is  \cite{Matyus2012}:
\begin{gather}
T_{IJ}=-\frac{1}{2}\sum_{i=1}^n\frac{\bra{\psi_I}\mx{\nabla}_{\mx{r}_i}\trs\left(\fett{M}\otimes\fett{1}_3\right)\mx{\nabla}_{\mx{r}_i}\ket{\psi_J}}{\left|\psi_I\right|\left|\psi_J\right|}\nonumber\\
=D^{3/4}\left(\frac{p_{\fett{u}_I,\fett{u}_I}}{q_{\fett{u}_I}}\right)^{K_I}\left(\frac{p_{\fett{u}_J,\fett{u}_J}}{q_{\fett{u}_J}}\right)^{K_J}\left(\frac{p_{\fett{u}_I,\fett{u}_J}}{\displaystyle\sqrt{q_{\fett{u}_I}q_{\fett{u}_J}}}\right)^L\sum_{m=0}^{\min(K_I,K_J)}\left(\frac{p_{\fett{u}_I,\fett{u}_J}p_{\fett{u}_I,\fett{u}_J}}{p_{\fett{u}_I,\fett{u}_I}p_{\fett{u}_J,\fett{u}_J}}\right)^m\nonumber\\
\times\left[R_{IJ}+(K_I-m)\frac{P_{\fett{u}_I,\fett{u}_I}}{p_{\fett{u}_I,\fett{u}_I}}+(K_J-m)\frac{P_{\fett{u}_J,\fett{u}_J}}{p_{\fett{u}_J,\fett{u}_J}}+(L+2m)\frac{P_{\fett{u}_I,\fett{u}_J}}{p_{\fett{u}_I,\fett{u}_J}}\right]H_{LK_IK_Jm}\label{eq:T}
\end{gather}
where $\fett{M}$ is a diagonal matrix with $M_{ii}=1/m_i$. The $H_{LK_IK_Jm}$ terms are precalculated factors
\begin{gather}
H_{LK_IK_Jm}=\frac{4^m(L+m+1)!}{(K_I-m)!(K_J-m)!m!(2L+2m+1)!}\times\frac{1}{\displaystyle\sqrt{F_{K_IL}F_{K_JL}}}\label{eq:8}
\end{gather}
in order to increase efficiency and ensure numerical stability with the terms
\begin{gather}
F_{KL}=\sum_{m=0}^K\frac{4^m(L+m+1)!}{(K-m)!(K-m)!m!(2L+2m+2)!}\label{eq:9}
\end{gather}
stemming from the quasi-normalisation
\begin{gather}
 \left|\psi_Z\right|=(\braket{\psi_Z}{\psi_Z})^{\frac{1}{2}}\quad \text{with} \quad Z\in\{I,J\}.
\end{gather}

Furthermore we introduced short hand notations for terms which we have to study in terms of their dependence on $c_A$. One term we have to study is
\begin{gather}
 D=\frac{\det(2\fett{A}_I)\det(2\fett{A}_J)}{\det(\fett{A}_{IJ})\det(\fett{A}_{IJ})}\label{eq:1}
\end{gather}
stemming from the origin centered Gaussian functions. Furthermore we have to analyse the terms depending on the global vectors
\begin{gather}
p_{\fett{u}_X,\fett{u}_Y}=\fett{u}_X^T\fett{A}_{IJ}^{-1}\fett{u}_Y\quad\text{with}\quad X,Y\in\{I,J\}\label{eq:3}\\
P_{\fett{u}_I,\fett{u}_I}=-\fett{u}_I^T\fett{A}_{IJ}^{-1}\fett{A}_J\fett{M}\fett{A}_J\fett{A}_{IJ}^{-1}\fett{u}_I\label{eq:5}\\
P_{\fett{u}_J,\fett{u}_J}=-\fett{u}_J^T\fett{A}_{IJ}^{-1}\fett{A}_I\fett{M}\fett{A}_I\fett{A}_{IJ}^{-1}\fett{u}_J\label{eq:6}\\
P_{\fett{u}_I,\fett{u}_J}=\fett{u}_I^T\fett{A}_{IJ}^{-1}\fett{A}_J\fett{M}\fett{A}_I\fett{A}_{IJ}^{-1}\fett{u}_J\label{eq:7},
\end{gather}
the terms from the quasi-normalisation
\begin{gather}
q_{\fett{u}_Z}=\frac{1}{2}\fett{u}_Z^T\fett{A}_Z^{-1}\fett{u}_Z \quad \text{with} \quad Z\in\{I,J\}\label{eq:2}
\end{gather}
and the term
\begin{gather}
 R_{IJ}=\frac{3}{2}\tr\left[\fett{A}_{IJ}^{-1}\fett{A}_J\fett{M}\fett{A}_I\right]\label{eq:4}
\end{gather}
which can be associated with the radial motion of the system \cite{Kinghorn1996}.

Here, we immediately see that the singularity of $\fett{A}_I$ and $\fett{A}_J$ would cause terms in Eqs.~(\ref{eq:1})--(\ref{eq:4}) to be not defined. The singularity is introduced if the $c_A=0$ selection is made to guarantee translation free expressions.

Since the potential energy terms in Eq.~(\ref{eq:Hop}) depend only on the inter-particle distances, we can choose any
$c_A>0$ value and evaluate the matrix elements without any problem and the resulting potential energy matrix elements are independent of the value of $c_A$. So, they are not discussed here any longer and are evaluated according to Ref. \cite{Matyus2012}.

\section{Identification and Elimination Strategy for the Translational Contamination}
\label{sec:TB}
In Section \ref{sec:ECG}, we noted that the ECGs take the same mathematical form, independent of whether we choose the LFCCs $\mx{r}$, or some $\mx{x}_\mr{TICM}$. This simple transformation behavior also transfers to the expressions of the integrals. Due to this property we can study the influence of $c_A$ on the different terms in Eqs.~(\ref{eq:1})--(\ref{eq:4}),
only the parameter matrices $\fett{A}_I,\fett{A}_J$  and $\fett{u}_I,\fett{u}_J$ corresponding to $\fett{r}$
have to be replaced by their according expressions in terms of $\fett{A}^{(x)}_I,\fett{A}^{(x)}_J$ and $\fett{u}^{(x)}_I,\fett{u}^{(x)}_J$ corresponding to $\fett{x}_\mr{TICM}$. The parameter matrices are related by the transformation given in Eqs.~(\ref{eq:amxtrans})--(\ref{eq:uvectrans}), and
$\fett{A}^{(x)}_I,\fett{A}^{(x)}_J$ and $\fett{u}^{(x)}_I,\fett{u}^{(x)}_J$ have block structure, Eq.~(\ref{eq:aublock}).

Firstly, let us consider the $R_{IJ}$ term of the kinetic energy matrix elements, Eq.~(\ref{eq:4}), explicitly.
We analyze the properties of $R_{IJ}$ with respect to $c_A$
using the TICC formalism:
\begin{align}
 R_{IJ}
 &=
 \frac{3}{2}\tr\left[\fett{A}_{IJ}^{-1}\fett{A}_J\fett{M}\fett{A}_I\right]
 \nonumber \\
 &=
 \frac{3}{2}\tr\left[(\fett{A}^{(x)}_{IJ})^{-1}\fett{A}^{(x)}_J\fett{U}_x\fett{M}\fett{U}_x^T\fett{A}^{(x)}_I\right].
\end{align}
and also exploit the block structure of the matrices:
\begin{gather}
 \fett{A}^{(x)}_{z}=\begin{bmatrix}
          \fett{\mathpzc{A}}^{(x)}_{z}&0\\
          0&c_{A}
               \end{bmatrix}\quad\text{with}\quad z\in\{I,J,IJ\}\quad\text{ and }\quad
\fett{U}_x\fett{M}\fett{U}_x^T=\begin{bmatrix}
          \fett{\mu}^{(x)}&0\\
          0&c_M
               \end{bmatrix}.
\end{gather}
The block structure of $\fett{U}_x\fett{M}\fett{U}_x^T$ follows from the conditions of translation invariance, Eqs.~(\ref{eq:ticond1})--(\ref{eq:ticond2}), for $\fett{x}$.
We can thus separate the $c_A$ dependent terms in $R_{IJ}$:
\begin{gather}
 R_{IJ}=R^{\mathrm{Int}}_{IJ}+\frac{3}{4}c_Ac_M=
 \frac{3}{2}\tr\left[%
   (\fett{\mathpzc{A}}^{(x)}_{IJ})^{-1}\fett{\mathpzc{A}}^{(x)}_{J}\fett{\mu}^{(x)}\fett{\mathpzc{A}}^{(x)}_{I} %
 \right]+\frac{3}{4}c_Ac_M
\end{gather}
The c$_M$ factor of the linear contribution is
\begin{gather}
 c_M
 =(\fett{U}_x\fett{M}\fett{U}_x^T)_{n+1,n+1}
 =\sum_{i=1}^{n+1}(\fett{U}_x)_{n+1,i}^2 / m_i
 =1/m_\mr{tot}.
\end{gather}
and thus with $c_M=1/m_\mr{tot}$
\begin{gather}
 R_{IJ}=R^{\mathrm{Int}}_{IJ}+\frac{3}{4}\frac{c_A}{m_\mr{tot}}
 \label{eq:cM}
\end{gather}
Next, we investigate the $c_A$ dependence of the remaining terms and factors, Eqs.~(\ref{eq:1})--(\ref{eq:2}), of the matrix representation of the kinetic energy in the LFCC formalism, Eq.~(\ref{eq:T}).
In Eq.~(\ref{eq:1}) we find that any contribution of $c_A$ cancels.
In Eqs.~(\ref{eq:2}) and (\ref{eq:3}) we see that $c_A$ only contributes if $c_u>0$. Since we require $c_u=0$, this contribution is eliminated.
Finally, in Eqs.~(\ref{eq:5})--(\ref{eq:7}) we find that the contribution of $c_A$ to the exponent matrices cancel. 
In short, only the $R_{IJ}$ term, Eq.~(\ref{eq:4}), has a non-vanishing (linear) $c_A$ dependence
in the kinetic energy matrix element, $T_{IJ}$.

Thus, if $c_A=0$ was chosen, the translational dependence vanishes, but the exponents matrices, $\fett{A}_I, \fett{A}_J$,
are singular without having an inverse. Thus, for an implementation in a computer program we can choose a
non-zero value for $c_A$, and eliminate the translational contribution explicitly by subtracting $3c_A/(4m_\mr{tot})$ from
the $R_{IJ}$ matrix element. This is a simple computational strategy which we are going to follow in the LFCC formalism.

We also note here that the direct variational optimisation of all parameters, including $c_A$ here, would be another option
that has been suggested already in the literature, e.g., in Ref.~\cite{Bochevarov2004}. From the theoretical details presented so far, we understand that the total energy with $c_A>0$ is always an upper bound to the total energy free of
the overall translation of the system, $E_\mr{tot}(c_A)\geq E_\mr{TI}$. 
But as we have shown for $c_A=0$ several parameter matrices incorporated in $E_\mr{tot}(c_A)$ are singular, so
in spite of the fact that the limit exists, the application in a computer program is problematic
($c_u=0$ is chosen throughout the discussion).

This explains our preference for the approach developed here, which releases the translation-free condition for the basis functions and corrects for the translational contamination in the kinetic energy explicitly for each basis function, $I=1,2,\ldots,N$. The details of our algorithm are:
\begin{enumerate}
 \item
   Generate, optimise or read in the $\alpha_{I,ij}$ values for $i=1,2,\ldots,n+1$, $j=i+1,i+2,\ldots,n+1$.
 \item
   Construct the elements of the exponent matrix in the LFCC formalism as \\
   \begin{align}
     (\mx{A}_I)_{ij}
     =
     -\alpha_{I,ij}(1-\delta_{ij})
     +\left(%
        \sum_{k=1,k\neq i}^{n+1} \alpha_{I,ik}
     \right)\delta_{ij}
     + c_A \frac{m_i}{m_\mr{tot}} \frac{m_i}{m_\mr{tot}}\nonumber
   \end{align}
   with $i,j=1,2,\ldots,n+1$ and $c_A>0$.
 \item
   Due to the $c_A>0$ choice the matrix $\mx{A}_I$ is non-singular, and $1/\det(\mx{A}_I)$ and $\mx{A}_I^{-1}$
   can be calculated. At the same time the total kinetic energy contains some translational contamination.
 \item
   The translational contamination is eliminated by replacing $R_{IJ}$, Eq.~(\ref{eq:4}),
   with $R_{IJ}-3c_A/(4m_\mr{tot})$ in the expression of the kinetic energy matrix element, $T_{IJ}$,
   Eq.~(\ref{eq:T}).
\end{enumerate}
Throughout this computational strategy for the elimination of the translational contamination in the LFCC formalism we have $c_u=0$, $c_A>0$.

\section{Numerical Examples}
\label{sec:NUM}
In this section we present numerical applications using the LFCC formalism. The appearance of the
translational contamination and its elimination according to the strategy described in Section~\ref{sec:TB} are demonstrated.

Our test cases are the lowest energy levels of the para-H$_2$ ($L=0,p=+1$) and the ortho-H$_2$ ($L=1,p=-1$)
molecules both in the singlet electronic state. These are the two lowest-energy rotational states of the hydrogen molecule. $L$ is the total spatial (orbital and rotational) quantum number
and $p$ is the parity.
The wave functions are obtained by a direct solution of the linear variational problem using
1500 basis functions with an optimised parametrisation taken from \cite{Matyus2012}.
Here we use the LFCC formalism exclusively, thus all parameters were transformed first to the
LFCC representation, according to Eqs.~(\ref{eq:amxtrans})--(\ref{eq:uvectrans}). The exponent matrix for each basis function
was constructed according to Eq.~(\ref{eq:Amxalpha}).
During this transformation we were free to choose $c_A$ to simulate different levels of translational contamination,
while $c_u=0$ was fixed. We used the same $c_A$ value for each basis function. Table~\ref{tab:1} collects the results of our numerical calculations.

\begin{table}[H]
 \caption{\label{tab:1}\small The appearance and elimination of the translational contamination of the total pre-Born--Oppenheimer energy using the laboratory-fixed-Cartesian-coordinate (LFCC) formalism. The ground-state energies, in \Eh , of the singlet, para-H$_2$ ($L = 0,p=+1$) and ortho-H$_2$ ($L=1,p=-1$) molecules are calculated using the basis function parameter set of Ref.~\cite{Matyus2012}. The corresponding translationally invariant energies are $E_{\mathrm{TI}}(L=0)  = -1.164\,025\,026$~\Eh\ and $E_{\mathrm{TI}}(L=1)=1.163\,485\,167$~\Eh, respectively.$^a$ Results corrected for the translational contamination of the LFCC formalism are indicated with the "corr." subscript.}
 \begin{center}
 \renewcommand{\baselinestretch}{1.0}
 \renewcommand{\arraystretch}{0.7}

  \begin{tabular}{cccccc}

\hline
\hline
$c_A^b$&$E_{\mathrm{LF}}^c$&$E_{\mathrm{LF,corr.}}^c$&$\bra{\Psi_{\mathrm{LF}}}\hat{T}_{\mathrm{LF}}\ket{\Psi_{\mathrm{LF}}}^c$&$\bra{\Psi_{\mathrm{LF}}}\hat{T}_{\mathrm{LF}}\ket{\Psi_{\mathrm{LF}}}_{\mathrm{corr.}}^c$&$\delta_{\mathrm{Tr.}}^c$\\
\hline
\multicolumn{6}{c}{para-H$_2$ $L=0$ }\\
\hline
0.01&-1.164022985&-1.164025026&1.164027045&1.164025004&0.000002041\\
0.10&-1.164004614&-1.164025026&1.164045416&1.164025004&0.000020412\\
0.50&-1.163922966&-1.164025026&1.164127064&1.164025004&0.000102060\\
1.00&-1.163820906&-1.164025026&1.164229124&1.164025004&0.000204120\\
2.00&-1.163616786&-1.164025026&1.164433245&1.164025004&0.000408240\\
\hline
\multicolumn{6}{c}{ortho-H$_2$ $L=1$}\\
\hline
0.01&-1.163483125&-1.163485167&1.163487203&1.163485161&0.000002041\\
0.10&-1.163464755&-1.163485167&1.163505573&1.163485161&0.000020412\\
0.50&-1.163383107&-1.163485167&1.163587222&1.163485161&0.000102060\\
1.00&-1.163281047&-1.163485167&1.163689282&1.163485161&0.000204120\\
2.00&-1.163076927&-1.163485167&1.163893402&1.163485161&0.000408240\\
\hline
\hline
  \end{tabular}\\
 \end{center}
\begin{flushleft}
\small
$^a$ The parametrisation of the internal basis set ($\mathcal{P}$) was taken from Ref.\ \cite{Matyus2012}. The number
of basis functions was $1\,500$, the order of the polynomial prefactors was $2K\in$ $[0, 20]$. For
the proton-electron mass ratio the value $m_\mr{p}$/$m_\mr{e}$ = $1\,836.152\,672\,47$
was used, and thus $m_\mr{tot} = 3\,674.305\,344\,94\ m_\mr{e}$.\\
$^b$ Free parameter of the explicitly correlated Gaussian basis functions expressed in LFCCs. The same $c_A$ value was used for each basis function in the basis set. The value $c_u=0$ was used throughout the calculationss. \\
$^c$ $E_{\mathrm{LF}}$ , $\Psi_{\mathrm{LF}}$ : Eigenvalue and eigenfunction of the full Hamiltonian expressed in LFCC using the basis function parameters $(\mathcal{P},c_A)$.\\
$\hat{T}_{\mathrm{LF}}$ : The kinetic energy operator expressed in LFCCs.\\
"corr." : Correction for the translational contamination in the LFCC formalism, as explained in Section~\ref{sec:Theo}.\\
$\delta_{\mathrm{Tr.}} = 3c_A/(4m_\mr{tot})$: Translational contamination of the kinetic energy, see Eq.~(\ref{eq:cM}). The corrected, translation-free value is obtained as
$E_{\mathrm{LF,corr.}} = E_{\mathrm{LF}} -\delta_{\mathrm{Tr.}}$.
\end{flushleft}
\end{table}

The first column of Table~\ref{tab:1} lists the five values of $c_A$ which have been studied. The next two columns provide the total energies of the system with and without translational contamination, respectively. The last two columns list the kinetic energy contribution with and without translational contamination, respectively. We observe that those energies, which contain translational contributions (columns 2 and 4) depend on the value of $c_A$ and that an increase of the value of $c_A$ causes an increase of the energy, according to Section~\ref{sec:TB}. We also note that the corrected energies (columns 3 and 5) are independent of $c_A$. Furthermore, we list the translational correction given in Eq.~(\ref{eq:cM}).
Hence, these results give a numerical confirmation that we have identified and eliminated
the translational contamination depending on the $c_A$ basis function parameter, while choosing $c_u=0$.

\section{Conclusion}
\label{sec:C}
In this work, we used for the first time laboratory-fixed Cartesian coordinates (LFCCs)
with an explicit correction strategy for the translational contamination
in pre-Born--Oppenheimer variational calculations with various angular momentum quantum numbers.
Our work was motivated by the inherent simplicity of the LFCCs and the corresponding operators.

Instead of transforming the coordinates we accounted for the
translational and rotational invariances of the isolated many-particle problem
by using an appropriate form and parametrisation of the basis functions in the variational procedure.
The basis functions were constructed using explicitly correlated Gaussians and the global vector representation,
and as an extension of our earlier work \cite{Matyus2012} we focused here on the usage of LFCCs and
the problem of translational invariance.

First of all, we observed that it is impossible to parametrise explicitly correlated Gaussian functions (ECGs) in such a way that the total system is at rest in LFCCs and at the same time the basis functions are square-integrable with a non-vanishing norm.

Fortunately, it was possible to devise a simple computational strategy to circumvent this problem.
So, for the sake of the stability of the numerical computations we released the translational constraint on
the basis functions, by choosing a non-zero value for the free parameter, $c_A$, in the LFCC parametrisation of the ECG exponents, and then explicitly corrected for the translational contamination in the kinetic energy integral expressions.
This correction term is a simple constant, which depends linearly on $c_A$. Its form was derived by considering
a few mathematical relationship of the formalism: a) the properties of the linear transformation between LFCCs and
translationally invariant and center-of-mass Cartesian coordinates (TICMCCs); b) the corresponding transformation
of the basis function parameter matrices; c) the fact that the parameter matrices are block diagonal in TICMCCs.

It was also shown that the uncorrected total energy $E_\mr{tot}(c_A)$ is an upper bound to the translation-free total (intrinsic) energy.
Thus, in principle, we could obtain this value by the variational optimisation of the LFCC parametrisation (implicitly
including the $c_A$ value in the optimisation). We prefer, however, our explicit treatment for the elimination of the
translational contamination, because in the $c_A=0$ limit the exponent matrix, $\mx{A}$, of each ECG would be singular,
and thus the numerical evaluation of $1/\det(\mx{A})$ and $\mx{A}^{-1}$ would be impossible.

Finally, to demonstrate the numerical applicability of our approach we calculated the lowest two rotational levels
of the singlet hydrogen molecule corresponding to the para and ortho proton spin states, respectively.

The presented LFCC formalism with the explicit translational contamination correction is an alternative but
equivalent to the traditional approaches using some set of translationally invariant Cartesian coordinates
with the Cartesian coordinates of the center of mass explicitly separated already in the Hamiltonian,
e.g. \cite{Matyus2012,Cafiero2003}. The simplicity of the LFCCs is an appealing choice
for the variational solution of the Schr\"{o}dinger equation with the non-relativistic Hamiltonian.  Furthermore, one can think of more complicated
operators for which the usage of the simplest possible coordinate representation and the avoidance of
any coordinate transformation is more than just a comfortable option.

\section{Acknowledgments}
This work has been supported by the Swiss National Science Foundation SNF (project 200020\_144458/1). EM.\ thanks the Hungarian Scientific Research Fund (OTKA, NK83583) for financial support.

\small
\providecommand{\refin}[1]{\\ \textbf{Referenced in:} #1}


\end{document}